\documentclass[showpacs,showkeys,aps,twocolumn]{revtex4}
\usepackage{color,graphicx}
\begin{document}

\title{The magnetization of PrFeAsO$_{0.60}$F$_{0.12}$ superconductor}

\author{D. Bhoi$^1$, P. Mandal$^1$, P. Choudhury$^2$, S. Dash$^3$ and A. Banerjee$^3$}

\address{$^1$Saha Institute of Nuclear Physics, 1/AF Bidhannagar,
Calcutta 700 064, India}
\address{$^2$Central Glass and Ceramic Research Institute, 196
Raja S. C. Mullick Road, Calcutta  700 032, India}
\address{$^3$UGC-DAE Consortium for Scientific Research, University Campus, Khandwa Road, Indore 452
017, India}
\date{\today}
\begin{abstract}

The magnetization of the PrFeAsO$_{0.60}$F$_{0.12}$ polycrystalline sample has been measured as functions of temperature and magnetic field $(H)$. The observed total magnetization is the sum of a superconducting irreversible magnetization ($M_s$) and a paramagnetic magnetization ($M_p$). Analysis of dc susceptibility $\chi(T)$ in the normal state shows that the paramagnetic component of magnetization comes from the Pr$^{+3}$ magnetic moments. The intragrain critical current density $(J_L)$ derived from the magnetization measurement is large. The $J_L(H)$ curve displays a second peak which shifts towards the high-field region with decreasing temperature. In the low-field region, a plateau up to a field $H^*$ followed by a power law $H^{-5/8}$ behavior of $J_L(H)$ is the characteristic of the strong pinning. A vortex phase diagram for the present superconductor has been obtained from the magnetization and resistivity data.
\end{abstract}
\maketitle
\newpage

\section{Introduction}

Flux pinning is one of the most important factors that governs the critical current density, $J_c$,  in a type-II superconductor. The critical current density is determined by the balance of two opposing forces acting on the magnetic flux lines: the pinning force due to the spatial variation of the condensation energy and the Lorentz force exerted by the transport current. Energy is dissipated whenever flux lines move. Traditionally, one distinguishes two regimes of dissipation: flux creep when the pinning force dominates \cite{ander} and flux flow when the Lorentz force dominates \cite{ybkim,tinkham}. A complete knowledge on the dynamics of the magnetic flux lines is, therefore, required in order to understand whether a system is potentially attractive candidate for technological applications. In cuprate superconductors, due to high anisotropy, short coherence length and high operating temperature, the pinning energy of the vortices is weak, leading to strong fluctuation and motion of the vortices. Because of the weak pinning, the energy dissipation caused by the thermally activated flux motion is present even at a finite current density, which limits the maximum value of $J_c$, and hence a variety of applications of the cuprates.

In this perspective, the pnictide superconductors are important for studying the vortex dynamics. The phenomenally high upper critical field $H_{c2}$ \cite{bhoi1,jaron,lee} and lower anisotropy observed in the pnictide superconductors indicate encouraging potential applications \cite{oka}. Among different families of pnictide superconductors, the growth of $A$Fe$_2$As$_2$ (122 system, where $A$ = Ba, Ca, Sr) single crystals is relatively easier \cite{milton,saha,prozo}. Eventually, the study of vortex dynamics in these systems has been extensively carried out \cite{prozo,kim,yang1,yama1,shen,kope}. The vortex dynamics in 122 single crystals has been analyzed using weak collective-pinning$-$collective flux-creep model. In the (Ba,K)Fe$_2$As$_2$ single crystal, a vortex-glass to vortex-liquid transition has been observed \cite{kim}. One of the prominent features in pnictide superconductors is the observation of a fishtail or second peak effect in the $M(H)$ curve \cite{prozo,kim,yang1,yama1,shen,kope}. Analysis of the magnetic relaxation data in Ba(Fe$_{0.93}$Co$_{0.07}$)$_2$As$_2$ single crystal shows that the field at which the fishtail magnetization is maximum, a crossover in the vortex dynamics from the collective to the plastic creep occurs \cite{prozo}. However, due to the scarcity of $Re$FeAsO$_{1-x}$F$_x$ (1111 system, where $Re$ = La, Ce, Pr, Nd, Sm, Gd, etc.) single crystals, vortex dynamics has been investigated mainly in polycrystalline samples. The study of vortex dynamics in polycrystalline SmFeAsO$_{0.9}$F$_{0.1}$ \cite{yang} and NdFeAsO$_{0.9}$F$_{0.1}$ \cite{prozo1} samples suggests a collective vortex pinning and creep mechanism in this system. In the present work, we have studied magnetization $M(T,H)$ in order to evaluate the critical state parameters related to the vortex dynamics and to construct the mixed-state phase diagram of the PrFeAsO$_{0.60}$F$_{0.12}$ superconductor.
\section{Sample characterization}
\begin{figure}[h]
  \includegraphics[width=0.5\textwidth]{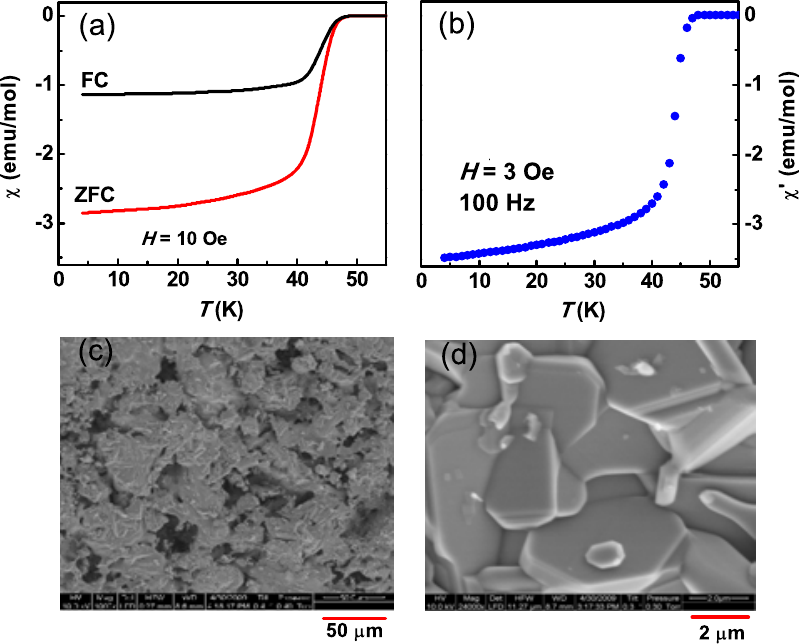}\\
  \caption{(a) Field-cooled and zero-field cooled dc susceptibilities at $H$ = 10 Oe and (b) real part of the ac susceptibility at $H$= 3 Oe and 100 Hz of the PrFeAsO$_{0.60}$F$_{0.12}$ sample. The slight paramagnetic susceptibility at the normal state has been subtracted from the above curves. (c) Low magnification and (d) high magnification scanning electron microscope images of the PrFeAsO$_{0.60}$F$_{0.12}$ sample. The high magnification image shows platelike crystallites of different sizes.}\label{Fig.1}
\end{figure}
We have synthesized oxygen deficient and fluorine-doped sample of nominal composition PrFeAsO$_{0.60}$F$_{0.12}$ by standard solid state reaction method. The details of the sample preparation have been discussed in our earlier reports \cite{bhoi}. The phase purity of the sample was determined by powder x-ray diffraction method with Cu K$\alpha$ radiation and details of the structural analysis are reported in Ref.\cite{bhoi}. For further characterization, we have measured the low-field magnetization of the sample using a SQUID magnetometer. The temperature dependence of zero-field-cooled (ZFC) and field-cooled (FC) dc magnetic susceptibilities at $H$ = 10 Oe are shown in figure 1(a). Compared to NdFeAsO$_{1-x}$F$_x$ \cite{prozo1,pissa} and other reported polycrystalline samples \cite{yama,yang,ren,yama2} the superconducting transition is rather sharp ($\Delta T_c \sim$ 4.5 K), indicative of better quality of the sample. Both the ZFC and FC susceptibilities start to deviate from the normal behavior below 48 K due to the appearance of a diamagnetic signal, which is close to the zero-resistance temperature $T_{c}^{0}$ \cite{bhoi}. At 4 K, the shielding and Meissner fractions are calculated to be 88\% and 35\% respectively, showing the bulk nature of superconductivity in the sample. We have also determined the shielding fraction of the sample $\sim$ 100\% at 4 K from the real part of the ac susceptibility curve [figure 1(b)]. In order to determine the average grain size in the sample, scanning electron microscope (SEM) (Quanta 200, FEG) images of the polished surface were obtained. SEM images of the polished surface of the PrFeAsO$_{0.60}$F$_{0.12}$ sample are shown in figure 1(c) and 1(d). The image in figure 1(c) reveals that the sample is porous and the conglomerate particle size varies between 20 and 50 $\mu$m. High magnification image of the particles shows plate-like crystallites with a size of 2-30 $\mu$m [figure 1(d)]. It may be mentioned that the average grain size in this sample is larger than that in LaFeAsO$_{0.89}$F$_{0.11}$, NdFeAsO$_{1-\delta}$ and SmFeAsO$_{0.85}$F$_{0.15}$ polycrystalline samples, where the grains are of the order of few hundreds of nanometer \cite{yama, moor,sena}. The energy dispersive x-ray (EDX) analysis was used to determine the chemical composition of the grains with different morphology. From the EDX spectra, we found that the oxygen content is about 0.75 and, the cations and fluorine contents are close to that of nominal composition. Examining the composition at several points on the surface of the sample, we did not find any local inhomogeneity. This result suggests that the grains are chemically homogeneous at least within the limit of SEM-EDX analysis. Hereafter, in the text we shall only mention the nominal composition of the sample. dc magnetization measurements at moderate and high fields were carried out using a Quantum Design superconducting quantum interference device with magnetic field up to 7 T. The magnetic hysteresis loops were measured using a 14 T vibrating sample magnetometer (Quantum Design). The magnetic hysteresis loops were measured at fixed temperatures with a constant field sweeping rate of 80 Oe s$^{-1}$.
\section{Results and Discussion}
\begin{figure}[t]
  \includegraphics[width=0.5\textwidth]{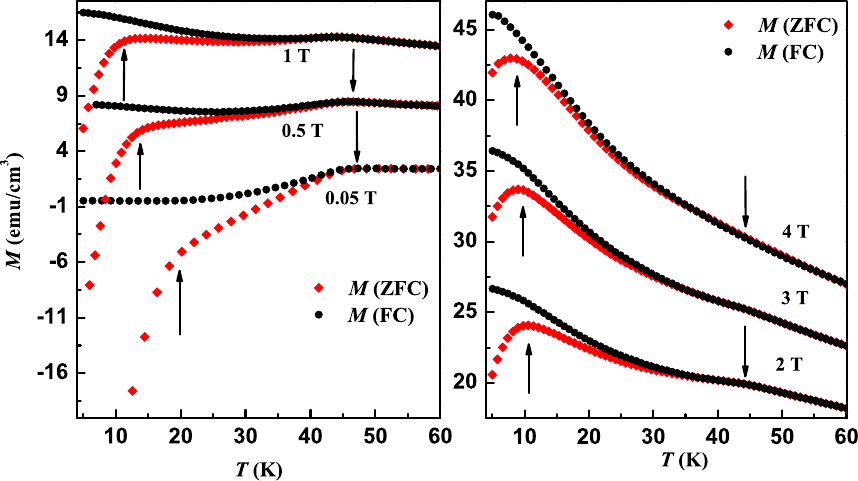}\\
  \caption{color online) Temperature dependence of the zero-field-cooled and field-cooled magnetization curves of the PrFeAsO$_{0.60}$F$_{0.12}$ sample under various applied magnetic fields. Left panel: for $H \leq$ 1 T. Right panel: for $H \geq$ 2 T. Upward arrows indicate the shoulder-like feature of the magnetization curve. Downward arrows indicate the superconducting onset transition temperature.}\label{Fig.2}
\end{figure}
Figure 2 shows the temperature dependence of ZFC magnetization ($M_{zfc}$) and FC magnetization ($M_{fc}$) of the PrFeAsO$_{0.60}$F$_{0.12}$ sample under different applied magnetic fields. For $H$ = 0.05 T, the $M(T)$ curve exhibits a usual superconducting behavior with the diamagnetic onset temperature $\sim$ 48 K. However, with the increase of magnetic field the difference between the ZFC and FC magnetizations reduces, and the magnetization becomes positive for fields above $H \simeq 1$ T as the paramagnetic component of the magnetization arising from the Pr ion overcomes the diamagnetic signal of the superconducting electrons. In contrast to the low-field data [figure 1(b)], the $M_{zfc}(T)$ curve for applied fields $H$$\geq$0.05 T displays a shoulder-like feature (shown in the figure with the upward arrow mark) indicating the onset of intergrain superconductivity which is expected in a granular superconductor. The granular behavior is related to the weak links of the grain boundaries. Weak links impose lower global (intergrain) critical current density in comparison to the intragrain critical current density. With the increase of magnetic field, the shoulder shifts towards the lower temperature region. This behavior is related to the magnetic flux penetration inside the individual grains and is determined by the crossover from intragrain to intergrain superconductivity. Similar granular behavior has also been observed in LaFeAsO$_{1-x}$F$_x$ \cite{yama}, SmFeAsO$_{1-x}$F$_x$ and NdFeAsO$_{1-x}$F$_x$ oxypnictides \cite{sena,yama2}.

In order to check that the paramagnetic component of the magnetization comes actually from the Pr$^{3+}$ ion, we have analyzed the temperature dependence of dc magnetic susceptibility measured at 500 Oe field in the normal state as shown in figure 3. The susceptibility curve can be fitted with the Curie-Weiss law, $\chi(T) = \chi(0) + C/(T+\theta)$ in the temperature region 90 K to 300 K, where $\chi(0)$ is the temperature independent Pauli susceptibility, $C$ = N$\mu_{eff}^2/3k_B$, the Curie constant, N is the number of Pr ions in the compound and $\theta$ is the Curie-Weiss temperature. The calculated effective moment $\mu_{eff}$ of the Pr$^{3+}$ ion is equal to 3.8 $\mu_B$ which is comparable with the derived magnetic moments 3.75 $\mu_B$ \cite{rotundu} and 3.64 $\mu_B$ \cite{mcguire} for the PrFeAsO samples. The small discrepancy from the effective moment of the free Pr$^{3+}$ ion (3.58 $\mu_B$) could be attributed to the possible crystalline electric field environment of the Pr$^{3+}$ ion \cite{rotundu}. Thus, we can ascribe the origin of paramagnetic background in the $M(T)$ data to the magnetic moment of the Pr ion. That is, the paramagnetic contribution to the total magnetization is intrinsic, not impurity related. In NdFeAsO$_{0.94}$F$_{0.06}$ sample too, large paramagnetic contribution to the $M(T,H)$ data has been observed because of the large effective magnetic moment of the Nd$^{3+}$ ion (3.62 $\mu_B$) \cite{tarantini}. The difference in the $M(T,H)$ curve from those of the La- and Sm-based Fe-pnictides is due to the large magnetic moment of the free Pr and Nd ions compared to those of La (= 0 $\mu_B$) and Sm ($\sim$ 0.8 $\mu_B$) ions.
\begin{figure}[t]
  \includegraphics[width=0.4\textwidth]{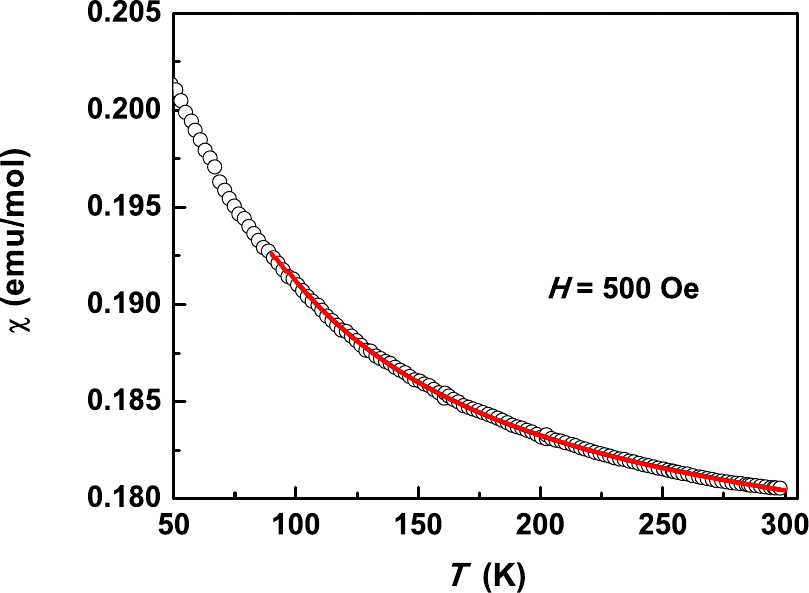}\\
  \caption{Temperature dependence of the dc susceptibility of PrFeAsO$_{0.60}$F$_{0.12}$ sample in the normal state measured at $H$= 500 Oe. The solid line is the result of the fit to Curie-Weiss law.}\label{Fig.3}
\end{figure}
Figure 4 presents the magnetization hysteresis $M(H)$ loops of the PrFeAsO$_{0.60}$F$_{0.12}$ sample at different temperatures in the field range -14 T $\leq$ H $\leq$ 14 T. One can clearly see that the superconducting hysteresis loop arises from the flux gradient produced by the pinning of flux lines. The observed $M(H)$ curve is the sum of a superconducting irreversible signal ($M_s$) and a paramagnetic component ($M_p$). In the presence of flux pinning, the sign of the superconducting magnetization depends on the direction of field sweep, thus resulting in a hysteresis loop between ascending and descending branches of the magnetization. The symmetric nature of the hysteresis curve suggests the bulk pinning of the vortices, not the surface barrier, dominates the magnetization of the sample. Even a 14 T magnetic field is not sufficient to achieve the closure of the width of the $M(H)$ loop at 25 K, indicating a large irreversible field of the superconductor. With increasing temperature, the irreversible diamagnetic loop shrinks and the paramagnetic component decreases significantly. Taking the mean value of the upper and lower hysteresis  branches of $M(H)$  as the paramagnetic component $M_p$ of the total magnetization, we deduce the superconducting magnetization due to the critical state of pinned vortices as $\Delta M$ = $(M^{+}- M^{-})$/2, where $M^{+}(M^{-})$ is the branch of the magnetization for ${dH}/{dt}<0$ (${dH}/{dt}>0$).
\begin{figure}
  \includegraphics[width=0.4\textwidth]{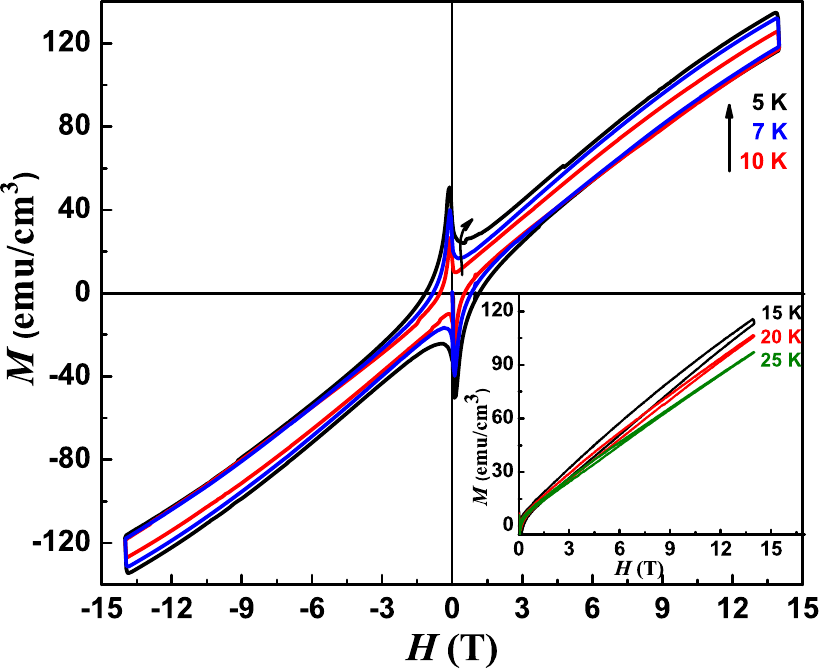}\\
  \caption{(color online) $M(H)$ loop of the PrFeAsO$_{0.60}$F$_{0.12}$ sample at 5, 7 and 10 K in the magnetic field range -14 T $\leq H \leq$ 14 T.Inset: $M(H)$ curve at 15, 20 and 25 K measured up to 14 T.}\label{Fig.4}
\end{figure}
In a polycrystalline sample, the gap $\Delta M$ in the magnetization loop can be split into an intergranular (global) part and an intragranular (local) part \cite{claim}, $\Delta M$ =  $\Delta M_G$ + $\Delta M_L$. Here, $\Delta M_G$ is the magnetic moment originating from the intergranular current density $J_G$ which flows across the whole sample dimension and is related to the grain-boundary Josephson junction current. The second term presents the sum over all the magnetic moments of the grains and is determined by the intragranular current density $J_L$ circulating within the grains. The intragranular current density is determined by the pinning of flux lines in the individual grains. A bulk polycrystalline superconductor can be described as a percolative network of weak links. The critical current is determined by the interconnection of individual superconducting regions, interconnections that take the form of normal or insulating tunnel barriers at grain or subgrain boundaries. The low value of the resulting transport critical current density is rapidly suppressed by an increase of the magnetic field or of the temperature. In low magnetic field region, the gap $\Delta M$ in magnetization loop is mainly caused by the intergranular current, but in the high-field region $\Delta M$ results largely due to the intragranular current. This has been confirmed from the magnetization loop measurements on the bulk sample and on powder prepared from the same bulk sample in cuprates and pnictides \cite{muller,chen}. Chen \emph{et. al.} \cite{chen} measured the magnetization on the bulk and powdered SmFeAsO$_{0.80}$F$_{0.20}$ samples at different temperatures and magnetic fields. In the high-field region, the magnetization of the bulk and powdered samples does not show any significant difference indicating that the gap in the magnetization loops is mainly due to the intragrain supercurrent instead of the current across grain boundaries. But, at zero field ($H$ = 0), the gap in the bulk sample is significantly larger than that for the powder sample, reflecting the contribution from the grain boundary supercurrent in the bulk sample. As the field increases from $H$ = 0 to a critical value, $\Delta M$ for the powder becomes slightly larger than that for the bulk due to the introduction of extra defects into the grains of the powder sample which act as additional pinning centers. The difference in magnetization between the two samples reduces rapidly with further increase of magnetic field. This indicates that the evaluation of the intragrain current from the gap in the magnetization loop of the powder sample will be overestimated in the low-field region. Excluding the low-field region close to $H$=0, we can assume that the gap $\Delta M$ in our bulk sample appears mainly due to the intragrain current density.

Assuming that the current is flowing within the grains, we can evaluate the intragrain current density by using the Bean formula $J_L = \frac{30\Delta M}{\langle R\rangle}$, where $\langle R\rangle$ is the average grain size \cite{bean}. Figure 5 shows the magnetic field dependence of the critical current density over the temperature range 5-35 K for an estimated average grain size of 5 $\mu$m as revealed by the SEM images. The estimated intragrain $J_L$ is 5$\times 10^5$ A cm$^{-2}$ at 5 K and 5 T which is in agreement with the result obtained from the magneto-optical imaging method (3$\times 10^5$ A cm$^{-2}$) in PrFeAsO$_{1-y}$ single crystal \cite{beek}. The value of intragrain $J_L$ for the present sample (5$\times10^5$ A cm$^{-2}$ at $H$ = 5 T and 5 K) is higher than that reported for other polycrystalline 1111 samples \cite{gao,wang} but comparable with the in-plane current density of 122 single crystal ($2.6\times 10^5$ A cm$^{-2}$ at 5 K and $H$ = 0) \cite{prozo}. The reported intragrain $J_L$ at $H$ =0 is $6 \times10^4$ A/cm$^{2}$ for SmFeAsO$_{0.7}$F$_{0.3}$, $2\times10^5$ A/cm$^{2}$ for SmFeAsO$_{0.65}$F$_{0.35}$ \cite{gao} and $2\times10^5 - 10^6$ A/cm$^{2}$ for NdFeAsO$_{0.82}$F$_{0.18}$ \cite{wang} polycrystalline samples.
\begin{figure}
  \includegraphics[width=0.4\textwidth]{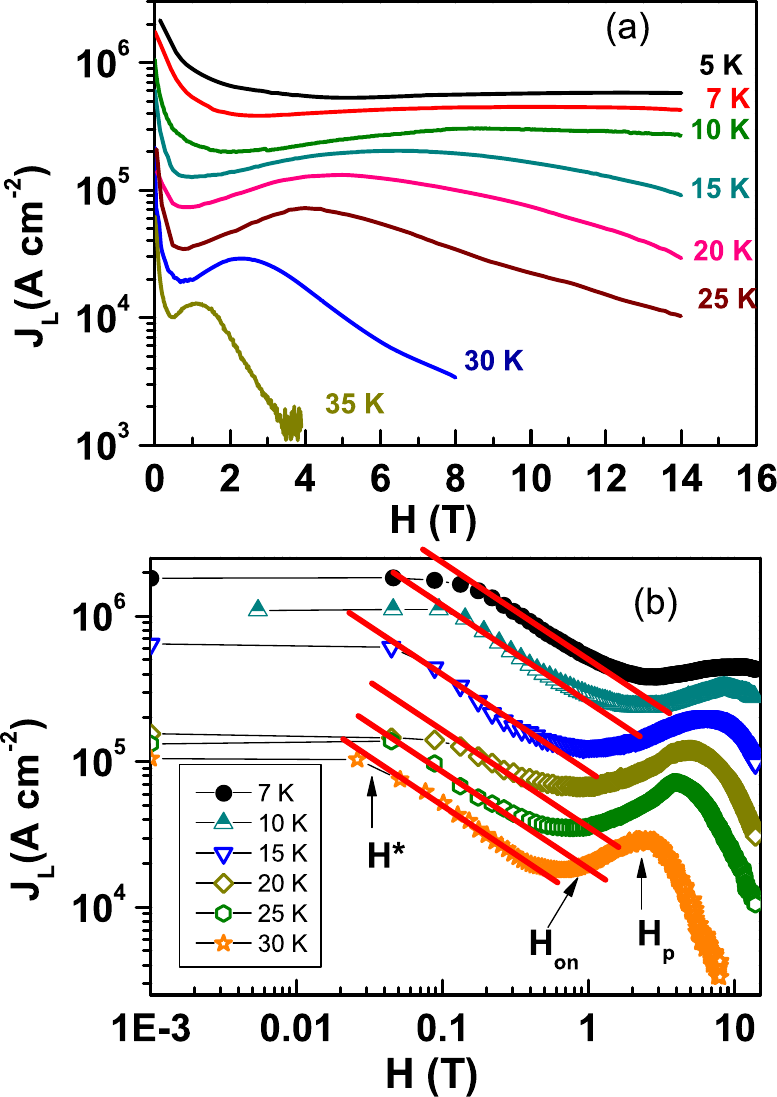}\\
  \caption{(color online) (a) Field dependence of the intragrain critical current density of PrFeAsO$_{0.60}$F$_{0.12}$ sample over the temperature range 5-35 K. (b) The $J_c(H)$ curve is plotted in double logarithmic scale for some selected temperatures. The solid lines show the power law $H^{-5/8}$ plot to the curve. The characteristic field $H^*$, $H_{on}$ and $H_p$ are indicated by arrows.}\label{Fig.5}
\end{figure}
In order to understand the pinning mechanism and to evaluate the parameters related to vortex dynamics, we have investigated the field dependence of the critical current density at different temperatures. A clear nonmonotonous field dependence of the intragrain critical current density is observed as shown in the figure 5(b) on a double logarithmic plot. [Figure 5(b) shows the field dependence of critical current density on a double logarithmic plot.] Up to a field $H^*$, the critical current density is almost independent of $H$ and then decreases rapidly with a power-law $H^{-5/8}$. In the intermediate field the critical current density saturates to a constant value $J^{SV}_c$. With further increase of the magnetic field above $H_{on}$, $J_c$ starts to increase and reaches the maximum value at $H_p$. In the plateau region, the single vortex pinning dominates. In the region between $H^*$ and $H_{on}$, the rapid decrease of critical current is due to the intervortex interaction. It has been claimed in Ref \cite{salem} that a crossover from single to collective pinning occurs above $H_{on}$. The constant value, $J^{SV}_c$, of the critical current density in the intermediate field region has been attributed to weak collective pinning of vortex lines by the small scale fluctuations of the local dopant atom density or oxygen vacancy \cite{beek}. The pinning mechanism is identified as being due to mean free path variations in the vortex core, where the dopant atoms acts as the effective quasiparticle scatterers. In this region, the critical current $J^{SV}_c$ is given as \cite{beek,beek1}
\begin{equation}
J^{SV}_c \simeq j_0 \left[\frac{0.1n_dD_v^4}{\varepsilon_\lambda\xi_{ab}}\left(\frac{\xi_0}{\xi_{ab}}\right)^2\right]^{2/3},
\end{equation}
where $j_0$ = 4$\epsilon_0$/$\sqrt3\Phi_0\xi_{ab}$ is the depairing current density, $n_d$ is the defect density, $D_v$ is the effective ion radius, $\varepsilon_{\lambda}$ = $\lambda_{ab}/\lambda_c$ is the low field anisotropy ratio of penetration depths, $\xi_{ab}$ is the inplane coherence length, $\epsilon_0$ = $\Phi_0^2$/$4\pi\mu_0\lambda_{ab}^2$ is the typical vortex energy scale, $\Phi_0$ is the flux quantum and $\xi_0 \simeq 1.35\xi(0)$ as the temperature independent Bardeen-Cooper-Schrieffer coherence length. The low temperature value of the $J^{SV}_c$ = 5$\times 10^5$ Acm$^{-2}$ can be reproduced if we assume $D_v \sim 1.35 \AA$ as the average ionic radius for oxygen and fluorine ion, with defect density $n_d\sim$ 3.5$\times 10^{27}$ m$^{-3}$. This value of $n_d$ corresponds to the 0.25 times oxygen vacancies in our sample. As our sample is simultaneously oxygen deficient and fluorine doped, this seems in consistent with the doping (25\% sum of oxygen vacancies and fluorine doping) in our sample. The calculated value of $n_d$ is 2.5 times larger than that of PrFeAsO$_{0.9}$ and NdFeAsO$_{0.9}$F$_{0.1}$ single crystals.

The field dependence of the critical current density, a plateau in the low-field followed by a power law decrease $J_c \propto H^{-5/8}$, is compatible with the theory of strong pinning \cite{beek1}. Similar field dependence of $J_c$ has also been observed in the case of PrFeAsO$_{1-y}$ single crystals \cite{beek}. In the presence of strong pins of size larger than the coherence length, it has been shown that \cite{beek1}
\begin{equation}
J_c(0)=\frac{\pi^{1/2}n_i^{1/2}j_0}{\varepsilon_{\lambda}}\left(\frac{f_{p,s}\xi_{ab}}{\epsilon_0}\right)^{3/2} (H < H^*),
\end{equation}
\begin{equation}
J_c(H)\approx\frac{2n_ij_0}{\varepsilon_{\lambda}^{5/4}\xi_{ab}^{1/2}}\left(\frac{f_{p,s}\xi_{ab}}{\epsilon_0}\right)^{9/4}\left(\frac{\Phi_0}{H}
\right)^{5/8} (H>H^*)
\end{equation}

where $n_i$ is the density of strong pins and $f_{p,s}$ is the elementary pinning force of a strong pin. The crossover field $H^*$=0.74$\varepsilon_{\lambda}^{-2}\Phi_0(n_i/\xi_{ab})^{4/5}(f_{p,s}\xi_{ab}/\epsilon_0)^{6/5}$ is determined as that above which the so called vortex trapping area of a single pin is limited by intervortex interactions. Using equation (2) and the power-law decrease of equation (3) we can determine $f_{p,s}$ from the ratio of [$dJ_c(H)/dH^{-5/8}$]/[$J_c(0)$]$^2$. Using the experimentally determined value of [$dJ_c(H)/dH^{-5/8}$]/[$J_c(0)$]$^2$ at 10 K and $\xi$ = 1 nm from our earlier work \cite{bhoi} along with the reported values of $\varepsilon_{\lambda}$ = 0.4, $j_0$ = 2$\times 10^{12}$ Am$^{-2}$ and $\epsilon_0$ = 3.2 $\times 10^{-12}$ Jm$^{-1}$ \cite{oka}, we have estimated the value of $f_{p,s} \approx$ 4$\times10^{-13}$ N for the present compound, which is two times larger than that for PrFeAsO$_{1-y}$ single crystal \cite{beek}. The density of pinning centers as estimated from $H^*$ is $n_i \approx$ 3$\times 10^{21}$ m$^{-3}$. van der Beek \emph{et al.} \cite{beek} attributed the source of strong pinning in PrFeAsO$_{1-y}$ and NdFeAsO$_{0.9}$F$_{0.1}$ single crystals to the extended (nm sized) pointlike inclusions or precipitates and to the spatial variations in the doping level on the scale of several dozen to 100 nm. They have also shown experimentally that in underdoped PrFeAsO, $T_c$ but also $J_c$ increase with an increasing density of oxygen vacancies, which corresponds to the introduction of more pinning centers. The larger values of pinning force, density of strong pins and higher $T_c$ in the present sample with higher doping level are consistent with their prediction.

The nonmonotonic behavior with a well pronounced second peak in magnetization represents the so-called "fishtail" effect. This phenomenon has been observed in conventional as well as high-temperature superconductors, but its origin is attributed to various mechanisms. The fishtail effect of comparable magnitude has also been observed recently in several iron arsenide superconductors \cite{prozo,kim,yang1,yama1,shen,kope,sena,beek,chen}. Most theoretical approaches agree that the temperature dependent field $H_p$, at which the second peak has its maximum, is related to vortex pinning and corresponds to a crossover between two different regimes of the vortex lattice. Various mechanisms have been proposed to explain the fishtail effect, foremost among them are a change in dynamics of the vortex lattice \cite{civa}, a change in flux creep behavior \cite{abu}, a change in its elastic properties \cite{blat} and a order-disorder transition of the vortex lattice due to the presence of topological defects \cite{kir,mik}. The peak effect in type-II superconductors was initially attributed by Pippard to the fact that the elastic constants of the vortex lattice vanish more rapidly than the pinning force as $H$ goes below $H_{c2}$ \cite{pippard}. Larkin and Ovchinnikov noted that this is exacerbated by the gradual softening of the vortex lattice because of the nonlocality of its tilt modulus \cite{larkin}. Both the above proposals were meant for the low-$T_c$ conventional superconductors where the peak effect occurs near the $H_{c2}(T)$ line. While for the present case the peak occurs for applied field well below the $H_{c2}(T)$ line which excludes the above scenario. The other models proposed to explain the second peak effect includes (i) a first order phase transition from an ordered "elastically pinned" low-field vortex phase, the so-called Bragg-glass, to a high-field disordered phase characterized by the presence of topological defects \cite{kir,mik} and (ii) a crossover from collective to plastic pinning. The former scenario has been verified in the case of high-temperature superconductors YBa$_2$Cu$_3$O$_{7-\delta}$ and Bi$_2$Sr$_2$CaCu$_2$O$_8$ \cite{koka,rass,beek3}, in the cubic superconductor (Ba,K)BiO$_3$ \cite{klein}, in NbSe$_2$ \cite{bhat}, in MgB$_2$ \cite{klein1} and recently, in PrFeAsO$_{1-y}$ and NdFeAsO$_{0.9}$F$_{0.1}$ single crystals \cite{beek}. However, discrepancies in the explanation of the second peak effect do exist in several cases. The peak effect in YBa$_2$Cu$_3$O$_{7-x}$ \cite{abu} and Nd$_{1.85}$Ce$_{0.15}$CuO$_{4-x}$ \cite{gill} has been ascribed to the crossover from collective to plastic pinning also. Majority of studies in 122 pnictide single crystals \cite{prozo,shen,salem} suggest that the peak effect is due to the crossover from collective to plastic pinning at $H$ = $H_p$.\\
\begin{figure}
  \includegraphics[width=0.4\textwidth]{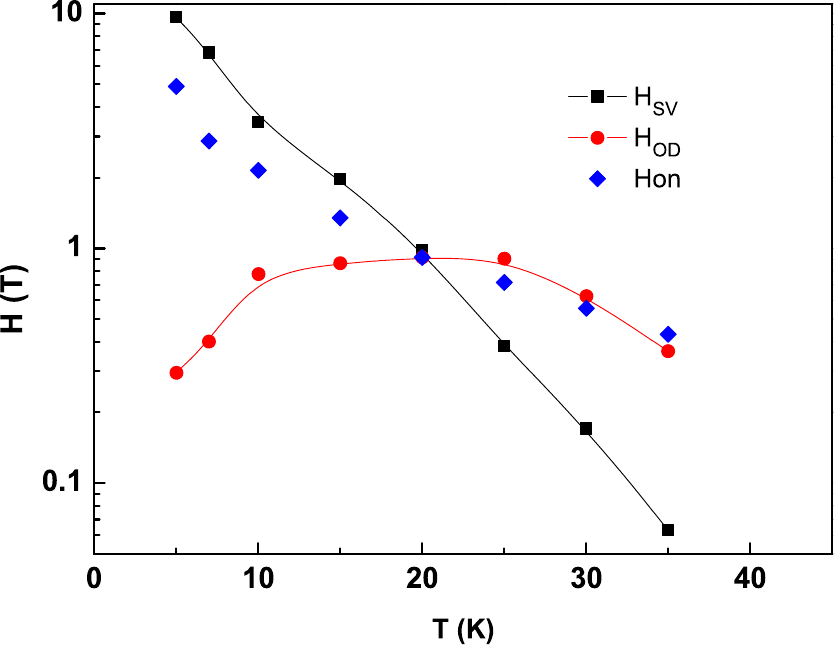}\\
  \caption{(color online) Temperature dependence of the calculated single-vortex to bundle pinning crossover field $H_{SV}$ and order-disorder field $H_{OD}$ at different temperatures with onset field $H_{on}$ data.}\label{Fig.6}
\end{figure}

In case of a crossover in vortex dynamics, the onset field $H_{on}$ should coincide with the single-vortex to bundle pinning crossover field $H_{SV}\sim 40H_{c2}\left(\frac{J^{SV}_c}{J_0}\right)$. However, if it is due to the order-disorder transition of the vortex lattice, then $H_{on}$ should coincide with the order-disorder transition $H_{OD}$ line. In the regime of single-vortex pining, the $H_{OD}$ is given by the following equation,
\begin{equation}
Ab_{SV}^{3/5}b_{OD}^{2/5}\left[1+\frac{F_T(t)}{b_{SV}^{1/2}(1-b_{OD})^{3/2}}\right]=2\pi c_L^2,
\end{equation}
where $b_{OD}\equiv H_{OD}/H_{c2}$, $b_{SV}\equiv H_{SV}/H_{c2}$, $c_L \sim$0.1 is the Lindemann number, A is a numerical constant, $t = T/T_c$, and $F_T(t) = 2t\left(\frac{Gi}{1-t^2}\right)^{1/2}$. Using $Gi$=0.01, $A$=4 and the experimental value of $J^{SV}_c$ at different temperatures, we have evaluated the $H_{SV}$ and $H_{OD}$ lines as shown in figure 6. Figure shows that $H_{on}$ data lie close to the $H_{OD}$ as compared to the $H_{SV}$ line which indicates the appearance of peak effect in this compound likely due to the order-disorder transition. A complete understanding of the second peak effect in pnictide superconductors needs further investigation.

\section{phase diagram}

In figure 7, we have drawn the $H$-$T$ phase diagram for the PrFeAsO$_{0.60}$F$_{0.12}$ superconductor as compiled from the magnetization and resistivity data \cite{bhoi1}. The magnetic upper critical field ($H_m$) determined from the superconducting onset transition of the $M$-$T$ curves under various applied fields is shown in the figure. For the sake of completeness, the resistive upper critical field $H_{c2}$ determined in our earlier measurement \cite{bhoi1} is also plotted in the phase diagram. The $H_m(T)$ curve appears closer to the resistive $H_{c2}(T)$ curve for 10$\%\rho_n$ criterion but far away from the $H_{c2}(T)$ curves for 90$\%\rho_n$ and 50$\%\rho_n$ criteria. It may be noted that the resistive transition reflects the net connectivity from one end to the other end of the sample through the superconducting domains while the magnetic measurement mainly reflects the bulk superconductivity  when most of the domains turn superconducting. In this respect, it is more plausible that the $H_m(T)$ line will appear closer to the 10$\%\rho_n$ $H_{c2}(T)$ line rather than 90$\%\rho_n$ or 50$\%\rho_n$ $H_{c2}(T)$ line.
\begin{figure}
  \includegraphics[width=0.4\textwidth]{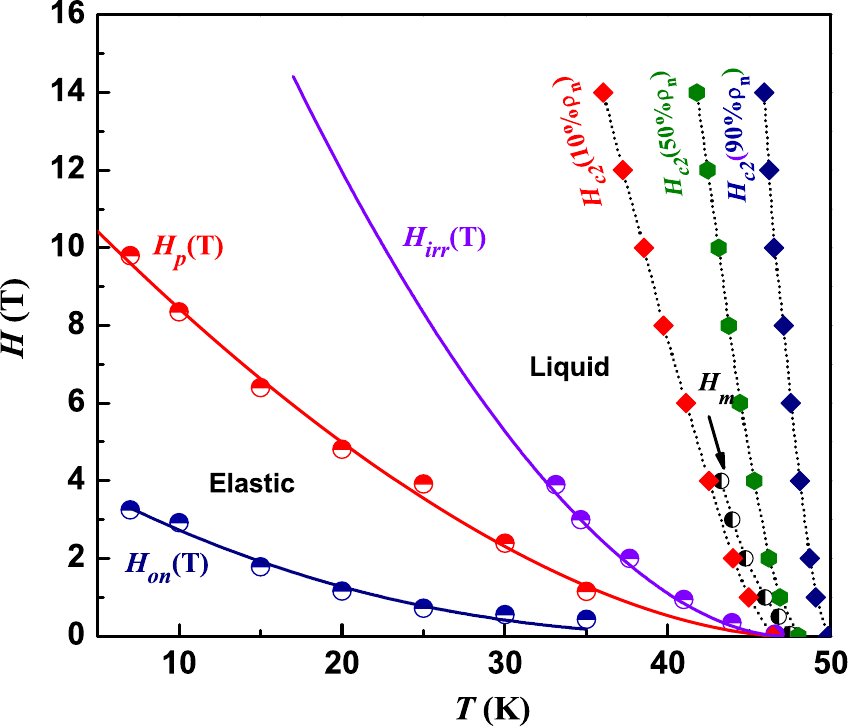}\\
  \caption{(color online) The vortex phase diagram of the PrFeAsO$_{0.60}$F$_{0.12}$ superconductor in the $H$-$T$ plane
  determined from the magnetization and resistivity data. The solid symbols describe $H_{c2}$ determined from 90$\%\rho_n$, 50$\%\rho_n$ and 10$\%\rho_n$ criteria. The solid lines are fit to the experimental $H_{on}$, $H_p$ and $H_{irr}$ data using Eq. (5) described in the text.}\label{Fig.7}
\end{figure}
A rough estimation of the irreversibility field $H_{irr}$ can be made from the $M(T)$ diagram where $M_{fc}(T)$ and $M_{zfc}(T)$ curves begin to diverge. Compared to the $H_m(T)$ line, the $H_{irr}(T)$ line shifts toward the lower temperature and lower field regions at a faster rate, indicating that there is a considerable gap between the upper critical field and the irreversibility field. Similar to cuprate superconductors, the $H_p(T)$ line is quite far from the $H_{c2}(T)$ line. In the phase diagram, we have also shown $H_{on}(T)$ and $H_{p}(T)$ curves. Normally, these characteristic fields, viz., $H_{irr}$, $H_{on}$ and $H_{p}$ follow a power-law expression of the form

\begin{equation}
H_x(T) = H_x(0)(1-T/T_c)^{n}.
\end{equation}

We observe that $H_{irr}(T)$, $H_{on}(T)$ and $H_p(T)$ data can be fitted well  by the above expression with exponent $n \sim$ 1.7, 2.4 and 1.6, respectively. The values of the prefactors are: $H_{irr}(0)$ = 31.9 T, $H_{on}(0)$ = 4.9 T and $H_p(0) = 12.5$ T. Such a temperature dependence of these characteristics fields has been observed in high-$T_c$ cuprates \cite{baily}, (Ba,K)BiO$_3$ \cite{blanc} and SmFeAsO$_{0.8}$F$_{0.2}$ \cite{chen}. The $H$-$T$ phase diagram can be broadly divided into three regions: (i) vortex liquid phase above $H_{irr}$, where magnetic flux lines are not pinned due to strong thermal fluctuations; (ii) the disordered phase between $H_p$ and $H_{irr}$, and (iii) the elastic ordered phase below $H_p$ lines.

\section{conclusion}
In summary, we have analyzed the temperature and field dependence of magnetization of the PrFeAsO$_{0.60}$F$_{0.12}$ superconductor. The magnetization of the compound can be understood as a sum of the superconducting irreversible signal, $M_s$ originating from the pinning flux lines and a paramagnetic component, $M_p$ arising from the magnetic moment of Pr$^{+3}$ ion. The critical current density exhibits a second peak at a magnetic field well below $H_{c2}(T)$ line. The critical current density is almost independent of $H$ up to a field $H^*$ and then decreases rapidly with a power law $H^{-5/8}$ which is the characteristic of strong pinning. The evaluated elementary pinning force, $f_{p,s}$ and the density of pinning centers, $n_i$ are larger than those for PrFeAsO$_{0.9}$ single crystal. This indicates that the oxygen deficiency is useful to increase the critical current density. The $H$-$T$ phase diagram for the present system has been obtained using the magnetization and resistivity data.

\section{Acknowledgement}

The authors would like to thank A. Pal and S. Banerjee for technical help during sample preparation and measurements. A. Banerjee would also like to thank DST, India for financial assistance for the 14 T vibrating sample magnetometer facility at CSR, Indore.

\end{document}